\documentclass[prl,showpacs,amsmath,amssymb]{revtex4}

\usepackage{amsmath}
\usepackage{amscd}
\usepackage{amsfonts}
\usepackage{amssymb}
\usepackage{mathrsfs}
\usepackage{fancybox} 
\usepackage{enumerate}
\usepackage{accents} % need to use \ring

\usepackage{graphicx}
\usepackage[dvips]{color}
\usepackage{ulem}

\if0
\makeatletter

\@addtoreset{equation}{section}
\makeatother

\usepackage{color}
\definecolor{blue1}{rgb}{0.15,0.15,0.50}
\usepackage[
debug,
dvipdfm,
colorlinks=true,
urlcolor=blue1,
anchorcolor=blue,
citecolor=cyan,
filecolor=blue,
linkcolor=blue1,
menucolor=blue,
pagecolor=blue,
linktocpage=true,
pageanchor=false,
]{hyperref}    
\fi

\def\tr{\mathrm{tr}}

\def\={\stackrel{\bullet}{=}}

\def\({\left(}
\def\){\right)}
\def\[{\left[}
\def\]{\right]}

\def \be {\begin{equation}}
\def \ee {\end{equation}}
\def \beqa {\begin{eqnarray}}
\def \eeqa {\end{eqnarray}}
\def \beal#1 {\begin{align}#1\end{align}}
\def \bes#1 {\begin{equation}\begin{split}#1\end{split}\end{equation}}
\def \nn {\notag\\}

\def\app#1{\stackrel{#1}{\ \approx\ }}

\begin{document}

\preprint{YITP-22-58\\
}

\title{
Noether's 1st theorem with local symmetries}
\author{
Sinya Aoki$^1$
}
\affiliation{ 
$^1$ Center for Gravitational Physics and Quantum Information,
Yukawa Institute for Theoretical Physics, Kyoto University,
 Kitashirakawa Oiwakecho, Sakyo-Ku, Kyoto 606-8502, Japan
 }

\email{saoki@yukawa.kyoto-u.ac.jp}

\begin{abstract}
Noether's 2nd theorem applied to a total system states that a global symmetry which is a part of local symmetries does not provide a physically meaningful conserved charge but it instead leads to off-shell constraints as a form of conserved currents.  
In this paper, we propose a general method to derive a matter conserved current associated with  a special global symmetry in the presence of local symmetries.
While currents derived from local symmetries of a  matter sector with a covariant background gauge field  are not conserved in general,
we show that the current associated with a special type of a global symmetry, called a hidden matter symmetry,   is on-shell conserved.   
We apply this derivation to a $U(1)$ gauge theory, general relativity and a non-abelian gauge theory. 
In general relativity, the associated conserved charge agrees with the one recently proposed from a different point of view.
\end{abstract}

\maketitle

%%%%%%%%%%%%%%%%%%%%%%%%%%%%%%%%%%%%%
\section{1. Introduction}
While the Noether's 1st theorem tells us how to define a conserved current if a theory is invariant under a global continuous transformation, 
 her 2nd theorem concludes that,  if a symmetry is local, 
 a conservation of a current is merely an identity or a constraint rather than a consequence from a dynamics of a system\cite{Noether:1918zz}.
For example, a trivially conserved current appear as $K ^\mu = \partial_\nu f^{\mu\nu}$ for an arbitrary anti-symmetric function $f^{\nu\mu}=-f^{\mu\nu}$.
 The Noether's 2nd theorem is applied also to a global symmetry which  is a part of some local symmetries\cite{Noether:1918zz,Deriglazov:2017biu,Aoki:2022gez}.
 This fact prevents us from defining a physically meaningful conserved energy in general relativity\cite{Noether:1918zz}.
 Indeed existing definitions of energy in general relativity correspond to conserved charges of the Noether's 2nd theorem,
 whose conservation is physically trivial\cite{Aoki:2022gez}.
 Furthermore it also brings us a question what is a conserved electric charge in electrodynamics, which has the local $U(1)$ gauge invariance.
 
 In this paper, we propose a general method to derive a matter conserved current associated with a special global symmetry which is a part of local symmetries.
 In Sec.~2,
 while a total system suffers from a Noether's 2nd theorem, a matter sector, which alone is still invariant under the local transformation  in the presence of a covariant background gauge field, is shown to have a conserved current of the Noether's 1st theorem for a special type of a global symmetry, which we call a hidden matter symmetry.
We apply the general method to three examples, a $U(1)$ gauge theory, general relativity and a non-abelian gauge theory in Sec.~3.
We show that an electric charge in the $U(1)$ gauge theory is conserved due to a constant $U(1)$ transformation of the matter sector.
In the case of general relativity, on the other hand, the vector for the global transformation giving a conserved current explicitly depends on the background metric like a Killing vector.  We show that our definition of a conserved charge in general relativity is nothing but the one recently proposed by
a collaboration including the present author without using symmetric argument\cite{Aoki:2020prb,Aoki:2020nzm}.
A background field dependence of the global transformation  plays an important role  in the non-abelian gauge theory  as well.
In Sec.~4, our conclusions and discussions are given. 
In particular, we stress that a difference between $U(1)$ gauge theory and others is a fact that local gauge transformations  are field independent only in the $U(1)$ gauge theory, which is related to an absence of non-linear terms in  the Lagrangian for gauge fields.

\section{2. Noether's 1st Theorem in a theory with local symmetries}
\label{sec:general}
We consider a theory with local symmetries, whose Lagrangian density with a parameter $\lambda$ is given by
\beqa
L_\lambda =L_G(g, g_{,\alpha},g_{,\alpha\beta}) +  \lambda\sum_n L_n(g,h_n,h_{n,\alpha}),
\eeqa
where $g=\{g_j\}$ is a gauge field, $h_n =\{h_n^j\}$ represents a matter field coupled to $g$ without its derivatives, 
and $g_{,\alpha} := \partial_\alpha g$,  etc.
Here $n$ labels a sector of matter fields such that $h_n$ and $h_m$ has no direct coupling between them if $n\not= m$.
We write 
$E^n\app{h_n} 0$ or $E^G \app{g} 0$ as an equation of motion (EOM) for $h_n$ or $g$, respectively,
where $\app{\Phi}$  means  that $\Phi$ is on-shell,  while $\approx$ means that all fields are on-shell.    
Explicitly we have
\beqa
E^n &:=&  {\partial L_n\over \partial h_n}-\partial_\alpha  \left({\partial L_n\over \partial h_{n,\alpha}}\right)\app{h_n} 0,\\
E^G&:=&  {\partial L\over \partial g}-\partial_\alpha  \left({\partial L_G\over \partial g_{,\alpha}}\right)
+\partial_\alpha \partial_\beta \left({\partial L_G\over \partial g_{,\alpha\beta}}\right)
\app{g} 0,
\eeqa

We assume that $L_G$ and $L_n$ are scalar density under infinitesimal local transformations generated by $\xi^a$ as
$\delta_\xi x^\alpha = X_a^\alpha(x) \xi^a(x)$,
\beqa
\delta_\xi g(x) &=& G_a (g)\xi^a(x) + G_a^\beta (g)\xi^a_{,\beta}(x),\nn
\delta_\xi h_n(x) &=& H_{n:a} (h_n,g)\xi^a(x) + H_{n:a}^\beta (h_n,g)\xi^a_{,\beta}(x),
\label{eq:gaugeTr}
\eeqa
where a summation over $a$ or $\beta$ is implicitly assumed.
It is useful to define $\bar\delta_\xi \Phi :=\delta_\xi\Phi -\Phi_{,\beta}\delta_\xi x^\beta$ with $\Phi=g,h_n$, 
since it commutes with derivatives as
$\bar\delta_\xi \Phi_{,\alpha} = \partial_\alpha \bar\delta_\xi \Phi$\cite{Utiyama:1984bc}.

\subsection{2-1. A total system and Noether's 2nd theorem} 
A local symmetry of an action integral for a total Lagrangian $L_\lambda$ leads to
\beqa
\int d^dx\left[ E^G \bar\delta_\xi g + \lambda \sum_n E^n \bar\delta_\xi h_n + \partial_\alpha N^\alpha_\lambda[\xi] \right] = 0,
\label{eq:1st_tot}
\eeqa
where
\beqa
N^\alpha_\lambda[\xi] &:=& N^\alpha_G[\xi] + \lambda \sum_n N^\alpha_n[\xi], \\
N^\alpha_G[\xi] &:=& {\partial L_G\over \partial g_{,\alpha}}\bar\delta_\xi g   
+ {\partial L_G\over \partial g_{,\alpha\beta}}\partial_\beta \bar\delta_\xi g 
- \partial_\beta \left({\partial L_G\over \partial g_{,\alpha\beta}}\right) \bar\delta_\xi g 
+ L_G \delta_\xi x^\alpha, \\
N^\alpha_n[\xi] &:=& {\partial L_n\over \partial h_{n,\alpha}}\bar\delta_\xi h_n + L_n \delta_\xi x^\alpha.
\eeqa
Thus the current $N^\alpha_\lambda[\xi]$ is on-shell conserved as
\beqa
 \partial_\alpha N^\alpha_\lambda[\xi]   &=& - E^G \bar\delta_\xi g - \lambda \sum_n E^n \bar\delta_\xi h_n
 \approx 0
\eeqa
for an arbitrary $\xi$ including a constant $\xi$. Therefore the Noether's 1st theorem for a constant $\xi$ seems to hold at first sight.
A fact that  $\partial_\alpha N^\alpha_\lambda[\xi] \approx 0$ for an arbitrary $\xi$, however, implies stronger relations which 
spoils a usefulness of the on-shell conserved current  $N^\alpha_\lambda[\xi]$, as shown below.

Using eq.~\eqref{eq:gaugeTr}, we rewrite eq.~\eqref{eq:1st_tot} as
\beqa
\int d^dx\, \left\{ O_{\lambda:a} \xi^a  +\partial_\alpha K^\alpha_{\lambda}[\xi]\right\} = 0,
\eeqa
where $O_{\lambda:a}$ is a $\xi$-independent function of fields $g$, $h_n$ and their derivatives, and 
\beqa
K^\alpha_{\lambda}[\xi] &:=& K^\alpha_{G}[\xi]+\lambda \sum_n K^\alpha_{n}[\xi]=N^\alpha_\lambda[\xi]+\left( E^G G^\alpha_a +\lambda \sum_n E^n H^\alpha_{n:a} \right)\xi^a, 
\label{eq:2nd_all}
\\
K^\alpha_{G}[\xi] &:=& N_G^\alpha[\xi] +\left( E^G -\lambda \sum_n {\partial L_n \over \partial g} \right) G^\alpha_a\xi^a, \\
K^\alpha_{n}[\xi] &:=& N_n^\alpha[\xi] + \left( {\partial L_n \over \partial g} G^\alpha_a + E^n H^\alpha_{n:a}\right) \xi^a.
\eeqa
The Noether's 2nd theorem tells us that $O_{\lambda:a} =0$ and $\partial_\alpha K^\alpha_{\lambda}[\xi]=0$ without using EOM\cite{Noether:1918zz,Utiyama:1984bc}.
Thus $K^\alpha_{\lambda}[\xi]$ is nothing but an explicit example of $K^\alpha =\partial_\beta f^{\alpha\beta}$ with $f^{\beta\alpha} = - f^{\alpha\beta}$, whose conservation is trivial.
Eq.~\eqref{eq:2nd_all} says that the conserved current $N_\lambda^\alpha[\xi]$ is equal to the trivial one up to EOMs as
\beqa
N^\alpha_\lambda[\xi] = K^\alpha_{\lambda}[\xi]  -  \left( E^G G^\alpha_a +\lambda \sum_n E^n H^\alpha_{n:a} \right)\xi^a \approx K^\alpha_{\lambda}[\xi],
\eeqa  
which is a conclusion from an application of the Noether's 2nd theorem to a global symmetry as a part of local symmetries\cite{Noether:1918zz}. 

\subsection{2-2. Noether's 1st theorem for a matter sector}
In order to escape from the Noether's 2nd theorem for the total system, we consider the matter action, whose invariance under the local transformation gives
\beqa
\int d^dx\, \left[{\partial L_n\over \partial g}\bar\delta_\xi g+ E^n \bar\delta_\xi h_n +\partial_\alpha N^\alpha_n[\xi] \right] = 0, 
\label{eq:1st}
\eeqa
where
\beqa
N^\alpha_n[\xi] &:=&  {\partial L_n\over \partial h_{n,\alpha}}\bar\delta_\xi h_n + L_n \delta_\xi x^\alpha.
\eeqa
Thus the current is NOT conserved for a general $\xi$ as
\beqa
\partial_\alpha N^\alpha_n[\xi] &=& - {\partial L_n\over \partial g}\bar\delta_\xi g - E^n \bar\delta_\xi h_n \app{h_n}  - {\partial L_n\over \partial g}\bar\delta_\xi g,
\eeqa
even if the EOM of the matter is employed.

If a vector $\zeta_n$ however satisfies 
\beqa
{\partial L_n\over \partial g}\bar\delta_{\zeta_n} g 
:= \sum_j {\partial L_n\over \partial g_j}\bar\delta_{\zeta_n} g_j 
\app{h_n} 0
\label{eq:cond_zeta}
\eeqa
for a fixed (background) $g$, which may or may not satisfy its EOM,
$N_n^\alpha[\zeta_n]$ is on-shell conserved as  
\beqa
\partial_\alpha N_n^\alpha[\zeta_n] \app{h_n} 0.
\eeqa
Thus  $N_n^\alpha[\zeta_n]$  is a physically meaningful conserved current of the Noether's 1st theorem for a global symmetry generated by a particular $\zeta_n(x)$,
which we call a hidden matter symmetry. 
Since eq.~\eqref{eq:cond_zeta} usually becomes one simple linear partial differential equation, it is easy to find a non-trivial solution in general,
as discussed in Ref.~\cite{Aoki:2020nzm} for the case of general relativity. In addition,
an explicit solution has been known  for time-dependent but spherically symmetric spacetimes in general relativity\cite{Kodama:1979vn}.
Since the EOM for $g$ is not required here, the conservation holds for an off-shell $g$ as well as an on-shell $g$, as long as $g$ is kept fixed in $L_n$. In this paper, the special vector is denoted as $\zeta_n(x)$ in order to avoid complicated expressions,
instead of using a more precise expression  $\zeta_n[g,h_n](x)$, which makes its dependence on $g$ and $h_n$ manifest. 

Furthermore, since $K_n^\alpha[\xi]$ is trivially conserved as a consequence of the Noether's 2nd theorem for a matter action,
we redefine the  current $N^\alpha_n[\xi]$ as
\beqa
J_n^\alpha[\xi] := N^\alpha_n[\xi]-K^\alpha_n[\xi] 
= -\left({\partial L_n\over \partial g} G^\alpha_a + E^n H_{n:a}^\alpha\right)\xi^a,~~
\label{eq:def_new}
\eeqa
which satisfies $\partial_\alpha J^\alpha_n[\xi] = \partial_\alpha N^\alpha_n[\xi]$.

Under \eqref{eq:cond_zeta},  the new current $J_n^\alpha[\zeta_n]$ is also conserved as
\beqa
\partial_\alpha J^\alpha_n[\zeta_n] &=&  - {\partial L_n\over \partial g}\bar\delta_{\zeta_n} g - E^n \bar\delta_{\zeta_n} h_n \app{h_n} 0,
\label{eq:conv_J}
\eeqa
where the conserved current $J_n^\alpha[\zeta_n]$ is given by
\beqa
J^\alpha_n[\zeta_n] &=& -\left({\partial L_n\over \partial g} G^\alpha_a + E^n H_{n:a}^\alpha\right)\zeta_n^a \app{h_n} - {\partial L_n\over \partial g} G^\alpha_a \zeta^a_n. \nn
\label{eq:def_J}
\eeqa
While $N^\alpha_n[\zeta_n]$ is related to the variation of  $L_n$ with respect to the matter field $h_n$,
the  on-shell $J^\alpha_n[\zeta_n]$ is determined by couplings between $g$ and $h_n$ in $L_n$.
Since $N^\alpha_n[\zeta_n]$ and  $J^\alpha_n[\zeta_n]$ are  equivalent up to the trivially conserved current $K^\alpha_n[\xi]$,
we focus our attention on  $J^\alpha_n[\zeta_n]$ in the remaining of this paper.
Thus, eqs.~\eqref{eq:def_J} and \eqref{eq:conv_J}, together with \eqref{eq:cond_zeta}, are main formulae used in this paper.

\section{3. Examples} 
\label{sec:example}
We apply the  method  in the previous section  to three examples, in order to construct a matter conserved charge for the hidden matter symmetry in each case.

\subsection{3-1. $U(1)$ gauge theory}
For a $U(1)$ gauge theory, we denote a gauge field $A_\mu$, and a matter field $\psi_{n j}$, which transform
under a $U(1)$ gauge symmetry by $\theta(x)$ as
\beqa
\delta_\theta A_\mu ={1\over e} \theta_{,\mu},  \ \delta_\theta \psi_{nj} = i q_{nj} \theta \psi_{nj},
\eeqa
where $e$ is a charge unit and $q_{nj} $ is a $U(1)$ representation of $\psi_{nj}$.

Eq.~\eqref{eq:def_new} leads to 
\beqa
J_n^\alpha[\theta] = -{1\over e} {\partial L_n\over \partial A_\alpha} \theta
=N_n^\alpha[\theta] = \sum_j i q_{nj} {\partial L_n\over \psi_{nj,\alpha}}\theta,
\eeqa
which satisfies
\beqa
\partial_\alpha J_n^\alpha[\theta] = 
-{1\over e}  {\partial L_n\over \partial A_\alpha} \partial_\alpha\theta
- i \sum_j  E^{nj} q_{nj} \psi_{nj}\theta, 
\eeqa
where $E^{nj}$ is an EOM for $\psi_{nj}$.
Therefore, for a constant $\theta(x) =\theta_0$, a generator of a hidden matter symmetry of this theory, 
$J_n^\alpha[\theta_0]$ becomes the on-shell conserved current from the Noether's 1st theorem for a global $U(1)$ symmetry generated by the constant 
$\theta_0$, and is equal to (a part of) the current  in the EOM for $A_\mu$ as
\beqa
{\partial L_G\over \partial A_\mu} -\partial_\alpha \left( {\partial L_G\over \partial A_{\mu,\alpha}}\right) = \lambda e \sum_n J^\mu_n[\theta_0=1].
\eeqa 

A fact that a global transformation generated by $\theta_0$ is independent of a gauge field $A_\mu$ is a reason why a conserved electric charge 
can be easily defined in electrodynamics at the classical level, and this property of the charge conservation can be faithfully carried over to QED.
Note that the electric charge is separately conserved in each $n$ even for the $U(1)$  gauge field $A_\mu$ to be dynamical, 
so that there appears no charge exchange between different sectors labeled by $n$. 

A distinction between matter fields and gauge fields becomes non-trivial if a $U(1)$ gauge symmetry is contained nontrivially in a part of a larger gauge symmetry. In the case of the electromagnetic $U(1)_{\rm em}$ symmetry  of the $U(1)_{\rm Y}\otimes SU(2)$ electroweak gauge symmetry, for example, 
a current $J_n^\alpha[\theta_0]$  defined from a matter action $L_n$ without including $W$ gauge bosons  is {\it not} conserved since
$\delta_{\theta_0} W_\mu^\pm \not=0$.  Thus we should include the action for $W$ gauge bosons into our definition of the matter action 
to ensure its invariance under the constant $U(1)_{\rm em}$ transformation.

\subsection{3-2. General relativity}
Under a general coordinate transformation by $\delta_\xi x^\alpha =\xi^\alpha$, the gravity (metric) and matter field, $g_{\mu\nu}$ and $\phi_n$,
transform as
\beqa
\delta_\xi g_{\mu\nu} &=&\bar \delta_\xi g_{\mu\nu} +g_{\mu\nu,\alpha}\xi^\alpha, \
\bar \delta_\xi g_{\mu\nu} = -(\nabla_\mu\xi_\nu +\nabla_\mu\xi_\mu), \nn
\delta_\xi \phi_n &=&H_n{}^\alpha{}_\beta \xi^\beta_{,\alpha}.
\eeqa
A $(k,l)$ tensor $T^A{}_B := T^{a_1\cdots a_k}{}_{b_1\cdots b_l}$ transforms as 
\beqa
\delta_\xi T^A{}_B = -{\cal L}_\xi  T^A{}_B + T^A{}_{B,\alpha}\xi^\alpha,
\eeqa
where ${\cal L}_\xi$ is the Lie derivative with the vector $\xi$.

For a matter sector $n$, eqs.~\eqref{eq:def_new} and \eqref{eq:conv_J} read 
\beqa
J_n^\alpha[\xi]&=& (2{\cal T}_n{}^\alpha{}_\beta - E^n H_n^\alpha{}_\beta )\xi^\beta \app{\phi_n} 2 {\cal T}_n^{\alpha\beta}\xi_\beta, \nn
\partial_\alpha J_n^\alpha[\xi]&=& {\cal T}_n^{\alpha\beta}\bar\delta_\xi g_{\alpha\beta} - E^n \bar\delta_\xi \phi_n \app{\phi_n}
{\cal T}_n^{\alpha\beta} \bar\delta_\xi g_{\alpha\beta},
\eeqa 
where $E^n$ is the EOM for the matter $\phi_n$, and
\beqa
{\cal T}_n^{\alpha\beta} =\sqrt{-g}  {T}_n^{\alpha\beta}:=\displaystyle {\partial L_n\over \partial g_{\alpha\beta}}
\eeqa 
is an symmetric energy momentum tensor (density),
which appears in the EOM for $g_{\mu\nu}$ as
\beqa
E^{g_{\mu\nu}}:={\partial L_G\over \partial g_{\mu\nu}} &-&\partial_\alpha \left({\partial L_G\over \partial g_{\mu\nu,\alpha}}\right)
+\partial_\alpha\partial_\beta \left({\partial L_G\over \partial g_{\mu\nu,\alpha\beta}}\right)
+\lambda \sum_n {\cal T}_n^{\mu\nu} \app{g_{\mu\nu}} 0 .
\label{eq:EOM_g}
\eeqa
On the other hand, $N^\alpha_n[\xi]$ consists of the canonical energy momentum tensor, 
and is related to $J^\alpha_n[\xi]$ made of the symmetric ${\cal T}_n^{\alpha\beta}$, by subtracting the trivially conserved current $K^\alpha_n[\xi]$.

If a vector $\zeta_n$ satisfies
\beqa
{\cal T}_n^{\mu\nu} [\nabla_\mu(\zeta_n)_\nu +\nabla_\nu(\zeta_n)_\mu ] \app{\phi_n} 0,
\label{eq:conv_cond}
\eeqa
$J^\alpha_n[\zeta^n] \app{\phi_n} 2 {\cal T}_n^{\alpha\beta}(\zeta_n)_\beta$ is an on-shell conserved current, and
 \beqa
 Q_n[\zeta^n]
 =2 \int_{\Sigma} d\Sigma_\alpha \, T_n^{\alpha\beta}(\zeta_n)_\beta,
  \label{eq:charge}
 \eeqa
 gives a conserved charge of  the Noether's 1st theorem in general relativity for the global hidden matter symmetry generated by  $\zeta_n(x)$,
 together with a symmetric energy momentum tensor $T_n^{\alpha\beta}$.
 The integration is performed over a $d-1$ dimensional space-like hyper-surface $\Sigma$ with the hyper-surface element $d\Sigma_\alpha$, and
 appropriate boundary conditions are assumed for the conservation of $Q_n[\zeta_n]$.
 The  condition \eqref{eq:conv_cond} for the vector $\zeta_n$  and
the corresponding conserved charge \eqref{eq:charge} in general relativity 
are identical to those previously proposed from a completely different argument\cite{Aoki:2020prb,Aoki:2020nzm}. 
Note that $Q_n[\zeta_n]$ is invariant under general coordinate transformations.

If a metric $g_{\mu\nu}$ allows a Killing vector $\xi_K$, 
  \eqref{eq:conv_cond} is satisfied by $\xi_K$, and thus $Q_n[\xi_K]$ is a conserved charge associated with the isometry of $g_{\mu\nu}$\cite{Aoki:2022gez,Aoki:2020prb}.
In this case, $\xi_K$ depends only on the metric $g_{\mu\nu}$  but is independent of matter fields $\phi_n$ for all $n$, so that
the symmetry generated by $\xi_K[g_{\mu\nu}]$ is a standard type of global symmetries for all matter sectors in the curved spacetime determined by
$g_{\mu\nu}$.
In particular, if the Killing vector $\xi_{K_0}$ is stationary, ${\cal E}_n:=Q_n[\xi_{K_0}]$ defines "energy" for the sector $n$ in general relativity\cite{Aoki:2022gez,Aoki:2020prb},
which is separately conserved in each sector.
Since the $\lambda\to 0$ limit implies a solution to \eqref{eq:EOM_g} with zero cosmological constant to be a Minkowski metric, 
${\cal E}_n$  with an appropriately normalized $\xi_{K_0}$ becomes the standard definition of energy in the flat spacetime as
\beqa
\lim_{\lambda\to 0} {\cal E}_n = \int_{x^0= {\rm fixed}} (d^{d-1}x)_0  \, T^{00} .
\eeqa

Even if a metric $g_{\mu\nu}$ has no Killing vector, we can always find a vector $\zeta_n$ which satisfies \eqref{eq:conv_cond}\cite{Aoki:2020nzm},
so that \eqref{eq:charge} defines an associated conserved charge, which is separately conserved in each sector even for the metric $g_{\mu\nu}$ to be dynamical.
If $\zeta_n$ is time-like, $Q_n[\zeta_n]$ defines energy or its generalization, which may be identified as  ``entropy" in some cases in general relativity\cite{Aoki:2020nzm}.
Note that $\zeta_n\not=\zeta_m$ in general if $n\not= m$. 

\subsection{3-3. Non-abelian gauge theory}
For a non-abelian gauge theory, the  local gauge transformation by $\theta=\theta^a t_a$ generates
\beqa
\delta_\theta  A_\mu &=& i[\theta, A_\mu]+{1\over g_s} \partial_\mu \theta, \  A_\mu := A_\mu^a t_a \nn
 \delta_\theta \psi_{nj} &=& i \theta^a t_a^{R_{nj}} \psi_{nj}, \  
\eeqa 
where $g_s$ is a coupling constant, $t_a$ is a generator in the fundamental representation of the gauge group  with an implicit  summation over $a$, while $t_a^{R_{nj}}$ is a generator in an irreducible representation  $R_{nj}$ for a matter $j$ in a sector $n$, so that $t_a^{R_{nj}}$  and $\psi_{nj}$ implicitly have gauge indices such that 
 $(t_a^{R_{nj}})^A{}_B$ and  $(\psi_{nj})^A$, respectively, where $A, B$ run from 1 to $\dim(R_{nj})$.

For a matter sector $n$, 
\eqref{eq:def_new} and \eqref{eq:conv_J} imply
\beqa
J_n^\alpha[\theta] &:=& -{1\over g_s} \tr ( T_n^\alpha \theta), \  (T_n^\alpha)^B{}_A:={\partial L_n\over \partial (A_\alpha)^A{}_B},\\
\partial_\alpha J^\alpha_n[\theta] &=& -  \tr (T_n^\alpha \delta_\theta A_\alpha) - \sum_j  (E^{nj} \delta_\theta \psi_{nj}),~~
\eeqa
where $E^{nj}$ is an EOM for $\psi_{nj}$, and
$T_n^\alpha$ appears in an EOM for $A_\mu$  as $\lambda \sum_n T_n^\alpha$.

If $\omega_n= \omega_n^a t_a$ (more precisely denoted as $w_n[A_\mu, \psi_{nj}]$) satisfies 
\beqa
\tr(T_n^\alpha \delta_{\omega_n} A_\alpha) = \tr\, T_n^\alpha \left\{ i[\omega_n,A_\alpha]  +{1\over g_s} \partial_\alpha \omega_n \right\} = 0,
\label{eq:inv_NA}
\eeqa
$J_n^\alpha[\omega_n]$ becomes  a conserved current for a hidden matter symmetry as 
\beqa
\partial_\alpha J^\alpha_n[\omega_n]\app{\psi_n} 0.
\eeqa
We then define a conserved non-abelian charge of the Noether's 1st theorem as
\beqa
Q_n[\omega_n] =\displaystyle  \int_{x^0={\rm fix}} (d^{d-1}x)_0\, J_n^0[\omega_n], 
\eeqa
which is invariant under  the gauge transformation $\Omega$ by
\beqa
A_\mu^\prime = \Omega\, A_\mu\, \Omega^\dagger -{1\over ig_s} \Omega\, \partial_\mu\Omega^\dagger,\quad
\omega_n^\prime =  \Omega\, \omega_n\, \Omega^\dagger, 
\eeqa
since $T_n^\alpha$ and $\delta_{\omega_n} A_\alpha$ are covariant as
\beqa
(T_n^\alpha)^\prime &=& \Omega \,T_n^\alpha\, \Omega^\dagger, \quad
\delta_{\omega_n^\prime} A_\alpha^\prime = i[\omega_n^\prime, A_\alpha^\prime] +{1\over g_s}\partial_\alpha \omega_n^\prime
=
\Omega (\delta_{\omega_n} A_\alpha) \Omega^\dagger,
\eeqa
so that $J_n^\alpha[\omega_n]$ and the condition \eqref{eq:inv_NA} are gauge invariant. 

Note again that the non-abelian gauge charge is separately conserved in each sector $n$ even for the non-abelian gauge field $A_\mu^a$ to be dynamical, so that there appears no non-abelian charge exchange
among different sectors. 

\section{4. Conclusions and discussions}
In this letter, we have proposed the general method to define the conserved current/charge for the global hidden matter symmetry  in the presence  of local symmetries,
which is then  applied to the $U(1)$ gauge theory, general relativity, and the non-abelian gauge theory.  
We have shown that the conserved electric charge in the electrodynamics  can be interpreted as this type of the conserved charge for the constant $U(1)$ transformation.
It also leads to a new type of conserved charges in general relativity and the non-abelian gauge theory. Indeed, 
the symmetric argument in our proposal theoretically explains 
an existence of a new conserved charge in general relativity recently proposed by different considerations\cite{Aoki:2020prb,Aoki:2020nzm}.

In the standard view, the electric field does not carry an electric charge because of an absence of non-linear terms,
while the gravitational field or the gluon carries energy or color, respectively,
due to non-linear interactions.
In our approach, however,
an absence or a presence of non-linear terms in the theory 
makes gauge transformations field independent or dependent, respectively,
so that 
the global hidden matter transformation, which leads to the conservation, is constant and field independent for the $U(1)$ theory but field dependent for others. 
Regardless of this difference,
we stress that the conserved charge in our method is carried only by matter fields for all cases. 
Of course, it would be better if we could define a physically meaningful conserved charge such as energy or non-abelian gauge charge including both matter and gauge contributions 
for a total system. Unfortunately one has failed  to find a global symmetry so far for such charges in the presence of a local symmetry.

We finally point out that our method can be extend to more general cases that the matter sector contains derivatives of gauge fields. 
For example, in the case that $L_n=L_n(g,g_{,\alpha}, h_n, h_{n,\alpha})$, 
\eqref{eq:def_J}, \eqref{eq:conv_J} and \eqref{eq:cond_zeta} still hold 
under a replacement that
\beqa
{\partial L_n\over \partial g}  \to {\partial L_n\over \partial g} - \partial_\beta\left( {\partial L_n\over \partial g_{,\beta}}\right).
\eeqa
\\

\section*{Acknowledgement}
This work is supported in part by the Grant-in-Aid of the Japanese Ministry of Education, Sciences and Technology, Sports and Culture (MEXT) for Scientific Research (Nos.~JP22H00129). 
The author thanks Prof. Yoshimasa Hidaka and Mr. Takumi Hayashi for useful discussions.

\end{document}